\documentclass[reqno,11pt,a4]{amsart}
\usepackage{amsmath,amssymb,tabularx,setspace,color}
\usepackage{pdflscape}
\newtheorem{theorem}{Theorem}[section]

\newtheorem{definition}{Definition}
\newtheorem{remark}{Remark}[section]

\usepackage{slashbox}
\usepackage{multirow}
\usepackage[bf,small,tableposition=top]{caption}
\usepackage{amsthm}
\usepackage{lineno}
\usepackage[figuresright]{rotating}
\usepackage{enumerate}
\usepackage[T1]{fontenc}
\usepackage[utf8]{inputenc}
\usepackage{graphics}
\usepackage{graphicx}
\usepackage{lscape}
\usepackage[dvipsnames]{xcolor}
\usepackage{mathrsfs}
\usepackage{hyperref}
\usepackage{textcomp}
\usepackage{listings}
\usepackage{amsmath}
\makeatletter
\renewcommand\section{\@startsection {section}{1}{\z@}%
                                   {-3.5ex \@plus -1ex \@minus -.2ex}%
                                   {2.3ex \@plus.2ex}%
                                   {\normalfont\large\bfseries}}
\makeatother

\newtheorem{Corollary}{Corollary}[section]

\begin{document}
\vspace{-1.9cm}
\doublespace
\title[]{U-Statistics for  Left Truncated and Right Censored Data}
\author[]%
{  S\lowercase{udheesh}, K. K.$^{\lowercase{a}\dag}$,  A\lowercase{njana}, S.$^{\lowercase{b}}$ \lowercase{and} X\lowercase{ie}, M.$^{\lowercase{c}}$\\
 $^{\lowercase{a}}$I\lowercase{ndian} S\lowercase{tatistical} I\lowercase{nstitute},
  C\lowercase{hennai}, I\lowercase{ndia,}\\
   $^{\lowercase{b}}$U\lowercase{niversity of} H\lowercase{yderabad}, H\lowercase{yderabad}, I\lowercase{ndia,}\\
  $^{\lowercase{c}}$C\lowercase{ity} U\lowercase{niversity of} H\lowercase{ong} K\lowercase{ong},  H\lowercase{ong} K\lowercase{ong}.
}
\thanks{{$^{\dag}$}{Corresponding author E-mail: \tt skkattu@isichennai.res.in}.}
\begin{abstract}
The analysis left truncated and right censored data is very common in survival  and reliability analysis. In lifetime studies patients often subject to left truncation in addition to right censoring.  For example, in bone marrow transplant studies based on International Bone Marrow Transplant Registry (IBMTR), the patients who die while waiting for the transplants will not be reported to the IBMTR.  In this paper, we develop novel U-statistics under   left truncation and right censoring. We prove the $\sqrt{n}$-consistency of the proposed U-statistics. We derive the asymptotic distribution of the U-statistics using counting process technique. As an application of the U-statistics, we develop a simple non-parametric test for testing the independence between time to failure and cause of failure in competing risks  when the observations are subject to left truncation and right censoring. The finite sample  performance of the proposed test is evaluated through Monte Carlo simulation study. Finally we illustrate our test procedure using lifetime data of transformers. 
\end{abstract}
\maketitle
\section{Introduction}
A common problem in practice is the presence of censoring where a competing event $C$ that
makes the variable of interest (lifetime) $X$ unobservable. That is, some of the lifetime may not be observed because of censoring.  For example, in the context of medical trials where patients often survive beyond the end of the trial period or are lost to follow-up for some reason.  In addition to censoring one may have encountered with left truncation, where we do not observe lifetime when $T<L$ where $L$ is the truncation random variable.  Left truncation occurs when the failure time of the subject under study is  included in the sample if the failure time is greater than the truncation time.

\par The present study  motivated by a real example of left truncated right censored data appeared in Hong et al. (2009). The data consist of the lifetimes of high voltage power transformers from an energy company in the US. There were approximately 15000 transformers and the company started recording the information about the transformers in 1980.  The data contain the information of transformers which are installed before or after 1980 but failed after 1980. In this paper, the we considered  the data till 2008. Thus lifetime of the transformers which are  still in service in 2008 is considered as right censored. Moreover, no information was available for the units which are installed and failed before 1980.  Hong et al. (2009), interested  in predicting the remaining lifetime of the transformers and the rate of failure of these transformers over time. They used a parametric model for explaining the distribution of lifetime of transformers. For the analyse of transformers data, Balakrishna and Mitra (2011, 2012, 2014) considered  different parametric models for lifetime distribution of transformers under left truncation and right censoring. Kundu et al. (2017) used the same data with further information about the cause of failures of transformers.  They considered the parametric analysis of the data in the presence of competing risks. The parametric models are more accurate only when we are able to correctly specify the underlying distribution. These facts motivated us to develop non-parametric inference for such data.

\par   Another example is the bone marrow transplant (BMT) studies using International Bone Marrow Transplant Registry (IBMTR). The patients who die while waiting for the transplants will not be reported to the IBMTR and those patient who lost to follow up is subject to random right censoring.   Hence it is important to study and develop methods to deal with  left truncated and right censored samples. We refer interested readers to Jiang et al. (2005), Klein   and Moeschberger (2006), Geskus (2011), Zhang et al. (2011),  Su and Wang (2012),  Vakulenko‐Lagun and  Mandel (2016),  Cortese et al.  (2017), Chen et al. (2017), Chen and Shen (2018), Efromovich and Chu  (2018), Hou et al. (2018), Jiang ae al. (2020) and Chen and Yi (2021)  and the references therein for some recent works based on left truncated and right censored data.

The theory of U-statistics has a major role in finding non-parametric  estimators of parameters of interest. Interested readers may refer to  Lee (1990) and  Kowalski  and  Tu (2007) for more discussion about the application of U-statistics in different fields.  Based on U-statistics, Jing et  al.  (2009) devolved jackknife empirical likelihood inference which has considerable attention recently. Due to  plethora of use cases in non-parametric inference it is desirable to develop U-statistics under different censoring scheme and truncations. In this scenario, an important concern is the reworking and extension of the procedures which exist for completely observed data. Using  inverse probability of censoring weighted (IPCW) approach, Datta et al. (2010) developed a right-censored version of U-statistics.  Satten et al. (2018) and Chen et al. (2019) discussed  comparing two distributions using two sample U-statistics in the presence of right censoring and confounding covariates. Motivated by these works, in this paper, an attempt is made to develop  U-statistics for left truncation and right censored data.

 The rest of the paper is organized as follows. In Section 2, we develop  novel U-statistics for left truncated and right censored  data.  We prove the consistency and asymptotic normality of the  proposed  U-statistics. We also obtain a consistent estimator of the asymptotic variance. In Section 3, making use of the U-statistics, we develop a new test  for testing the independence between cause of failure and failure time in  competing risks under left truncated and right censored  data set up.  The finite sample performance of the test is evaluated through Monte Carlo simulation study. The proposed method is illustrated using lifetime data of transformers. Concluding remarks along with some open problems are given in Section 4.
 \vspace{-0.18in}
\section{Proposed  U-statistics}
\vspace{-0.15in}
Suppose  $X$, $C$ and $L$ denote the failure time, censoring time and truncation time random variables respectively. Let $F(.)$, $G(.)$ and $H(.)$ be the distribution functions corresponding to the random variables $X$, $C$ and $L$, respectively. Also denote $\bar F(x)=1-F(x)$, $\bar G(x)=1-G(x)$ and $\bar H(x)=1-H(x)$. We assume the conditional independence  between $C$ and $L$ given $X$.  Under right censoring and left truncation, observed sample consists of $n$ independent and identically distributed observations $(T_i\epsilon_i, \delta_i)$ from $(T\epsilon, \delta)$, where $T= \min(X,C)$, $\delta=I(X<C)$ and $\epsilon=I(T>L)$. Here $\delta$ is the censoring indicator, while   $\epsilon$ is used to specify the truncation. Clearly $T_i$ is observed for $i$-th subject only when $T_i>L_i$.

We start by defining U-statistics for complete (uncensored) data. Let $X_1, ...,X_n $ be a random sample of size $n$ from $F$.    Let $h(X_1,...,X_m)$ be a symmetric kernel of degree $m$ with the property $E\big(h(X_1,...,X_m)\big)=\theta$, where $\theta$ is real.
The U-statistics with symmetric kernel $h$ is defined as
\begin{equation*}\label{ustat}
U=\frac{1}{\binom{n}{m}}\sum_{1\le i_1<\ldots<i_m\le n}h(X_{i_{1}}, ...,X_{i_{m}}),
\end{equation*}where we use ${i_1}, ...,{i_m}$ to indicates $m$ integers chosen from $(1,\ldots,n)$.
U-statistics  are widely used for finding  estimators of  several statistical functionals as well as for developing non-parametric tests as the asymptotic properties of them are well studied. Lehmann (1951) proved the  strong consistency   of U-statistics.  Moreover, the asymptotic distribution of $U$ is Gaussian with mean $\theta$ and variance $m^2\sigma^2$ (see Theorem 1, Chapter 3
of Lee (1980)) where $\sigma^2=Var\big(E\big(h(X_1,...,X_m)|X_1\big)\big)$.

Next we define U-statistics for left truncated and right censored (LTRC) data.  We use IPCW approach to define  U-statistics. To find the weight used in IPCW approach we consider
\begin{eqnarray}\label{identity}
E(\delta\epsilon T )&&=E(\delta \epsilon X )\nonumber\\
 &&=E\big(X E(\epsilon | X)E(\delta| X) \big)\nonumber\\
  &&=E\big(X P(C>X|X)P(X>L|X)\big)\nonumber\\
 &&=E(XP(C>X>L|X)).
\end{eqnarray}	
The first identity follows from the fact that $T=X$ when $\delta=1$. We also  use the fact that $L$ and $C$ are conditionally independent given $X$. In view of equation (\ref{identity}), we consider a weight function $\frac{\delta \epsilon }{P(L<T<C)}$ for defining  the U-statistics when the sample contain   left truncated and right censored observations.

We define U-statistics for LTRC data as
\begin{equation}\label{us}
U_m= \frac{1}{\binom{n}{m}} \sum_{1\le i_1<\ldots<i_m\le n}\frac{h(T_{i1},...,T_{im}) \prod_{l\in \underline{i}} \delta_l\epsilon_l}{\prod_{l\in \underline{i}} P(L_l<T_l<C_l)}.
\end{equation}
provided $P(L_i<T_i<C_i)>0$, for each $i$, with probability one.   Here,  the notation $l\in \underline{i}$ is used to indicate that $l$ is one of the integers $\{i_1,i_2\ldots,i_m\}$ chosen from $(1,\ldots,n)$.
For $m=1$ and $m=2$ we have
\begin{equation*}
U_1= \frac{1}{n} \sum_{i=1}^{n}\frac{h(T_{i})  \delta_i\epsilon_i}{P(L_i<T_i<C_i)}
\end{equation*}and
\begin{equation*}
U_2= \frac{2}{n(n-1)} \sum_{i=1}^{n-1}\sum_{j=i+1}^{n}\frac{h(T_{i},T_{j})  \delta_i\delta_j\epsilon_i\epsilon_j }{P(L_i<T_i<C_i)P(L_j<T_j<C_j)}.
\end{equation*}
Under the conditional independence of $C$ and $L$ given $T$, using equation (\ref{identity}), it can be easily verified that the U-statistics defined in equation (\ref{us}) is an unbiased estimator of $\theta$. Since  $U_m$ is a U-statistic with kernel $\frac{h(T_{i1},...,T_{im}) \prod_{l\in \underline{i}} \delta_l\epsilon_l}{\prod_{l\in \underline{i}} P(L_l<T_l<C_l)}$ it  is a consistent estimator of $\theta$ and has asymptotic normal distribution.

As $P(L_i<T_i<C_i)$ appeared in the equation (\ref{us})  is not known, we need to estimate it. We estimate it by   $\widehat K_c(T_i)$ and is given in equation (\ref{kmest}) below. Hence we define IPCW U-statistics under left truncation and   right censoring as
\begin{equation*}\label{ustat}
\widehat U_m=\frac{1}{\binom{n}{m}}  \sum_{1\le i_1<\ldots<i_m\le n}\frac{h(T_{i1},T_{i2},...,T_{im}) \prod_{l\in \underline{i}} \delta_l\epsilon_l}{\prod_{l\in \underline{i}} \widehat K_c(T_l)}.
\end{equation*}

\noindent Next we study the asymptotic properties of $\widehat U_m$. First we establish the consistency of  $ \widehat U_m$. The proof of the following theorem is given in Appendix.
\begin{theorem}
  Assume $E|h(T_1,T_2,...,T_m)|<\infty$. As $n\rightarrow \infty$, $ \widehat U_m$ converges in probability to $\theta.$
\end{theorem}

Next, we obtain the asymptotic distribution of  $\widehat U_m$.  For $i=1,2,...,n$, we define $N_i(t)=I(T_i \le t, \delta_i=1)$ and  $N_i^c(t)=I(T_i \le t, \delta_i=0)$  as the counting process corresponding to failure time and censoring time, respectively.   Also, denote $N(t)=\sum_{i=1}^{n} N_{i}(t)$, $N^c(t)=\sum_{i=1}^{n} N_{i}^{c}(t)$.  We define risk indicator as  $Y_i(t)=I(T_i \ge t \ge L_i)$ and  $Y(t)=\sum_{i=1}^{n}Y_i(t)$. Note that the risk set $Y(t)$ at $t$ contains the subjects entered the study before $t$ and are still under study at $t$.
 Clearly $N_{i}^c(t)$ is local sub-martingale with appropriate filtration $\mathbb{F}_t$. The martingale associated with the censoring counting  process with filtration $\mathbb{F}_t$ is given by
 \begin{equation}
 \label{3.1.1}
 M_i^c(t)=N_{i}^c(t)-\int_0^t Y_i (u) \lambda_c(u)du,\qquad i=1,2\ldots,n,
 \end{equation}
 where $\lambda_c(.)$ is the hazard   function corresponding to the censoring variable $C$ under left truncation.  The cumulative hazard function is given by $\Lambda(t)=\int_{0}^{t}\lambda(t)dt.$ We also denote $M^c(t)=\sum_{i=1}^{n} M_i^c(t)$.

 Also, we define the sub-distribution function of $T_1$ corresponding to $\delta_1=1$ and $\epsilon_1=1$ as
 \begin{equation}\label{subdist} S(x)=P(T_1 \le x, \delta_1\epsilon_1=1).\end{equation} Let
\begin{equation}\label{3.1.1.c}w(t)=\int_{0}^{\infty}\frac{h_1(x)}{P(L_1\le x\le C_1)}I(x>t)dS(x),\end{equation}
where $h_{1}(x)=E(h((T_1,\delta_1),\ldots ,(T_m,\delta_m))|(T_1,\delta_1)=(x,\delta_1))$. Also we denote  $y(t)=P(T_1 \ge t \ge L_1)$. An estimator of the survival function of censoring variable  $C$ under left truncation, denoted by  $\widehat K_{c}$, is given by
\begin{equation}\label{kmest}
  \widehat K_{c}(\tau)=\prod_{t\le\tau}\Big(1-\frac{dN^{c}(t)}{Y(t)}\Big).
\end{equation}As an analog to the  Nelson-Aalen estimator,  the estimator for cumulative hazard function for $C$ under left truncation is defined as
\begin{equation}\label{naest}
  \widehat \Lambda_{c}(\tau)=\int_{0}^{\tau}\frac{dN^{c}(t)}{Y(t)}.
\end{equation}In both the definitions given in equations (\ref{kmest}) and (\ref{naest}) we assume $Y(t)$ is non-zero with probability one.  The relationship between $ \widehat K_{c}(\tau)$ and  $ \widehat \Lambda_{c}(\tau)$ is given by
\begin{equation}\label{kmrel}
  \widehat K_{c}(\tau)=\exp[-\widehat \Lambda_{c}(\tau)].
\end{equation}  Next we state the assumptions needed to prove the asymptotic distributions.
\begin{enumerate}[{C1:}]
   \item $E(h((T_1,\delta_1),\ldots ,(T_m,\delta_m)))<\infty$,
  \item $\int\frac{h_{1}^2(x)}{\widehat K_{c}^2(x)}dS(x) <\infty$,
  \item $\int \frac{w^2(x)\lambda_c(x)}{y(x)}dx <\infty$.
\end{enumerate}
Now we find the asymptotic distribution of $\widehat U_m$ and the proof is given in  Appendix.
\begin{theorem}\label{thmad}
  Under the conditions $C1$-$C3$,  as $n\rightarrow \infty$,   $\sqrt{n}(\widehat U_m-\theta)$ converges in distribution to  Gaussian random variable with mean zero and variance $ m^2\sigma_{c}^2$, where  $ \sigma_{c}^2$ is given by
\begin{equation}
\label{3.1.2}
\sigma_{c}^2=\sigma_{1}^{2}+\sigma_{2}^{2}
\end{equation}with
\begin{equation*}
  \sigma_{1}^{2}=Var\Big(\frac{h_1(X)\delta_1\epsilon_1}{K_c(X)}\Big)
\end{equation*}
and
\begin{equation}\label{limvar}
  \sigma_{2}^{2}= \int_{0}^{\infty}\frac{w^2(x) d{\Lambda}_c(x)}{y(x)}.
\end{equation}
\end{theorem}

Next we find a consistent estimator of the asymptotic variance $ \sigma_{c}^2$. Using the re-weighting principle an estimator of  $h_1(x)$ is given by
 \begin{equation}\label{hath1}
\widehat{h}_1(x)=\frac{1}{n^m}\sum_{1\le i_2<\ldots<i_m\le n} \frac{h(x,T_{i_{2}},...,T_{i_{m}}) \delta_{i_{2}}\ldots  \delta_{i_{m}}\epsilon_{i_{2}}\ldots\epsilon_{i_{n}}}{\widehat K_c(T_{i_{2}})\ldots\widehat K_c(T_{i_{m}})}.
 \end{equation}A consistent estimator  $\sigma_{1}^2$ is given by
\begin{equation*}
\widehat{\sigma}_{1}^{2}=\frac{1}{n-1}\sum_{i=1}^{n}(V_{i}-\bar V)^2,
\end{equation*}
where
\begin{equation*}\label{36}
V_{i}=\frac{\widehat h_1(T_i)\delta_i\epsilon_i}{\widehat K_c(T_i)}\qquad\text{and}\qquad   \bar V=\frac{1}{n}\sum_{i=1}^{n}V_{i}.
\end{equation*}
 Using equations (\ref{3.1.1.c}) and (\ref{hath1}), we find  an estimator of $w(x)$  as
\begin{eqnarray}\label{hatomega}
   \widehat{w}(x)&=&\nonumber \int_{0}^{\infty}\frac{\widehat{h}_1(z)I(z>x)}{\widehat{K}_c(z)}d\widehat{S}(z)\\
   &=& \frac{1}{n}\sum_{i=1}^{n}\frac{\widehat{h}_1(T_i)I(T_i>x)\delta_i\epsilon_i}{\widehat{K}_c(T_i)}.
\end{eqnarray}We can  estimate $y(x)$ by $\frac{Y(x)}{n}$. Hence using (\ref{naest}), from equation (\ref{limvar}) we obtain an estimator of  $\sigma_{2}^2$ as
\begin{equation*}
\widehat \sigma_{2}^2=\sum_{i=1}^{n}\frac{n\widehat w^2(T_i)(1-\delta_i)} {Y^2(T_i) }.
\end{equation*}
Hence an estimator of the asymptotic variance $\sigma_{c}^2$ is given by
\begin{equation*}
\widehat\sigma_{c}^2=\frac{1}{n-1}\sum_{i=1}^{n}(V_{i}-\bar V)^2+\sum_{i=1}^{n}\frac{\widehat w^2(T_i)(1-\delta_i)} {Y^2(T_i)/n }.
\end{equation*}

\section{Test for competing risks}
In classical survival studies, the study subjects are at risk of one terminal event. However in many survival studies, the failure (death) of an individual may be due to one
of $k$ (say) causes. In such situations, each unit under study is exposed to $k$ causes of failure, but its failure can be due to exactly one of
these causes of failure. The data that arise from such contexts are known as competing risks data.   In literature, competing risks are either considered through a latent failure time approach or through bivariate random pair $(X,J)$, where $X$ is the failure time of the unit and $J\in\{1,2,...,k\}$ is the corresponding cause of failure. In our study we restrict to the case $k=2$. For  modeling and  analysis of competing risks data, one may refer to Kalbfeisch and Prentice (2002), Lawless (2003) and
Crowder (2012).

The joint distribution of $(X,J)$ is specified by the sub-distribution functions $$F_r(t)=P(X\le t,J=r),\quad r=1,2.$$ Then $F(x)=F_1(x)+F_2(x)$.  We define $P_r=P(J=r)$, $r=1,2$ with $P_1+P_2=1$.  Also, define sub-survival function $S_r(t)=P(T > t,J=r)$, $r=1,2$ and $S(t)=P(T> t)=S_1(t)+S_2(t)$.

In the analysis of competing risks data using observable random vector $(X,J)$, testing the dependence between $X$ and $J$ is very important. If $X$ and $J$ are independent, then $F_r(t)=P(J=r)F(t)$ and hence one can study $X$ and $J$ separately. Accordingly, the testing problem reduces to test whether $P(J=1)=P(J=2)=\frac{1}{2}$ and the bivariate problem is converted to the problem involving only $J$. We refer interested readers to  Dewan et al. (2004) and Anjana et al. (2019) and the reference therein for more details on testing the independence between $X$ and $J$.  Anjana et al. (2019) discussed how to incorporate the right censored observation in their methodology.  For analysing the  transformer data discussed above we are interested in developing a test for testing  independent between $X$ and $J$ under left truncated  right censored situation.

\subsection{Test statistics}
 Next, we discuss the testing problem in the presence of right censoring  and left truncation.  Suppose $T$, $X$, $C$ and $L$  are the random variables as defined in Section 2  and $J$ denote the random variable corresponds to the cause of failure.     Under right censoring and left truncation, we observe the competing risks data as $(T\epsilon,\delta,J\delta)$, where   $\epsilon$ and $\delta$ are defined as in Section 2.  Let $(T_i\epsilon_i,\delta_i,J_i\delta_i)$, $i=1,2,...,n$ be  independent  copies of $(T\epsilon,\delta,J\delta)$.
 On the basis of the observed data, we are interested to test the null  hypothesis
\begin{equation}
\label{2.1}
H_0: T \,\text{and}\, J\, \text{are independent}\nonumber
\end{equation}
against the alternative hypothesis
\begin{equation*}
  H_1: T\, \text{and}\, J\, \text{are not independent}.
\end{equation*}
To detect the departure from the null hypothesis $H_{0}$ towards the alternative hypothesis $H_1$, we consider a measure $\Delta$ given by

\begin{eqnarray*}\label{measure}
\Delta &=&\int_{0}^{\infty}(S_2(t_1)S_1(t_2)-S_2(t_2)S_1(t_1))dF_1(t)dF_2(t)\nonumber \\
&=& P(T_1>T_2>T_3>T_4, J_1=2,J_3=1)\nonumber\\&&\nonumber\qquad-P(T_1>T_2>T_3>T_4, J_1=1,J_3=2).
\end{eqnarray*}
\noindent It can be easily verified that $\Delta$ is zero under $H_0$ and positive under $H_1$. We use the U-statistics defined in Section 2 to find the test statistic $\widehat\Delta$. Define the kernel $\psi_{c}^{*}$
\begin{small}
\begin{equation*}
\label{3}
\psi_{c}^{*}((T_1,J_1),(T_2,J_2),(T_3,J_3)(T_4,J_4))=\begin{cases}
1\, &\text{if}  \begin{cases}T_1>T_2>T_3>T_4,J_1=1,J_3=2 \end{cases} \\
-1 & \text{if} \begin{cases}T_1>T_2>T_3>T_4,J_1=2,J_3=1.\end{cases}
\end{cases}
\end{equation*}
\end{small}
Hence the test statistics is given by
\begin{equation*}
\label{teststat}
\widehat\Delta=\frac{1}{\dbinom{n}{4}} \sum\limits_{1\le i<j<l<k\le n}\frac{\psi_c((T_i,J_i),(T_j,J_j),(T_l,J_l)(T_k,J_k))\delta_i\delta_j\delta_l\delta_k\epsilon_i\epsilon_j\epsilon_l\epsilon_k} {\widehat K_c(T_{i})\widehat K_c(T_{j})\widehat K_c(T_{l})\widehat K_c(T_{k})},
\end{equation*}
where $\psi_c$ is the symmetric version corresponding to $\psi_{c}^{*}$.
Test procedure is to reject the null hypothesis $H_0$ against the  alternative hypothesis $H_1$ for large values of $\widehat\Delta$. We obtain a   critical region of the test based on the asymptotic distribution of $\widehat\Delta$.
 In the next theorem, we find the limiting distribution of $\widehat \Delta $. The proof of the following theorem follows from  Theorem \ref{thmad}.
 \begin{theorem}Let \begin{small}$\psi_{1}^c (x)=E\big(\psi_c((T_1,J_1),(T_2,J_2),(T_3,J_3)(T_4,J_4))|T_1=x,J_1=j\big)$\end{small}.
Assume $E(\psi_c(T_1,T_2,T_3, T_4, J_1, J_2,J_3,J_4))<\infty$,  $\int\frac{(\psi_{1}^{c }(t))^2}{  y^2(t)}dS (t) <\infty$ and $\int \frac{w^2(t)}{y (t)}\lambda_c(t)dt <\infty$.  As $n\rightarrow\infty$,  $\sqrt{n} (\widehat\Delta-\Delta)$ is distributed as Gaussian with mean 0 and variance $16\sigma_{1c}^2$, where $\sigma_{1c}^2$ is given by
\begin{equation*}
\label{3.1.2}
\sigma_{1c}^2= Var\left(\frac{\psi_{1}^{c }(T_1)\epsilon_1\delta_1}{\widehat K_c(T_1)}\right)+\int \frac{w^2(t)}{y (t)}\lambda_c(t)dt.
\end{equation*}
 \end{theorem}
 A consistent  estimator of $\sigma_{1c}^2$ is given  by
\begin{equation*}\label{ecvar}
\widehat \sigma_{1c}^2=\frac{1}{n-1}\sum_{i=1}^{n}(V_{i}-\bar V)^2+\sum_{i=1}^{n}\frac{n\widehat w^2(T_i)(1-\delta_i)}{Y^2(T_i)},
\end{equation*}
where
\begin{equation*}\label{36}
V_{i}=\frac{\widehat\psi_{1}^{c}(T_i)\epsilon_i\delta_i}{\widehat K_c(T_i)}\qquad\text{and}\qquad\bar V=\frac{1}{n}\sum_{i=1}^{n}V_{i} 
\end{equation*}and the estimators  $\widehat\psi_{1}^{c}(x)$ and $\widehat{w}(x)$  can be obtained using the expressions (\ref{hath1}) and (\ref{hatomega}), respectively by considering the kernel $\psi_c(.)$. 

Using the asymptotic distribution obtained above, we obtain a critical region of the test. For large values of $n$, we reject the null hypothesis $H_0$ in favour of $H_1$ if
\begin{equation*}
 \frac{ \sqrt{n}\widehat{\Delta}_c}{\widehat{\sigma}_{1c}}>Z_{\alpha},
\end{equation*}where $Z_{\alpha}$ is the upper $\alpha$-percentile points of the standard normal distribution.

\subsection{Simulation study}
Next, we report the results of  Monte Carlo simulation study  carried out to evaluate the performance of the proposed test procedure.   The simulations are carried out using R software and repeated 10000 times.
\begin{table}[h]
	\caption{Empirical type 1 error of the test}

\begin{tabular}{|l|l|ll|ll|}
	\hline
	$(a,P_1)$& $n$&\multicolumn{2}{l|}{  20\% Censored}& \multicolumn{2}{l|}{  40\% Censored}\\
	\cline{3-6}
	&&  5 \% level &1 \% level&  5 \% level  &1 \% level\\
	\hline
	\multirow{6}[10]{*}{$(1,0.45)$}
	&	50&0.0541& 0.0145&0.0552 &0.0154\\
	&	75&0.0524 &0.0130 &0.0538 &0.0132\\
	&	100&0.0512&0.0119 &0.0521 &0.0121\\
	&	150&0.0506 &0.0111 & 0.0513&0.0115\\
	&  200&0.0503&0.0105 &0.0511 &0.0110 \\
	\hline
	\multirow{6}[10]{*}{$(1,0.48)$}
	&	50&0.0642& 0.0162&0.0661 &0.0164\\
	&	75&0.0622 &0.0155 &0.0632 &0.0160\\
	&	100&0.0592&0.0140 &0.0603 &0.0151\\
	&	150&0.0566 &0.0132 & 0.0573&0.0136\\
	&  200&0.0523&0.0115 &0.0530 &0.0119 \\
			\hline
		
\end{tabular}%
\label{ta1}%
\end{table}%

Lifetime random variable $X$ is generated from standard  exponential distribution. Censoring variable $C$ is generated from exponential distribution with parameter $\gamma$, where $\gamma$ is chosen such a way that the sample contains  desired percentage of censored observations, that is, $P(X>C)=p$, $0<p<1$. In the present study we considered  two situations with    $20 \%$ and $40 \%$ of  the observations are censored. The truncated variable $L$ is generated from exponential distribution with $\lambda$, where $\lambda$ satisfy $P(L>X)=0.2$, which guaranteed $20\%$ observations are truncated.

\par For generating the random samples from competing risks with two causes of failure, we consider the parametric family of sub-distribution functions given by (Dewan and Kulathinal, 2009).
\begin{align}
\label{4.1}
F_1(t)=P_1 F^{a}(t)\quad \text{and}\quad F_2(t)=F(t)-F_1(t),\nonumber
\end{align}
where  $1 \le a \le 2$ and $0<P_1<1$. If  $a = 1$ , then $T$ and $J$ are independent.

\par We computed the empirical type 1 error at 5\% and 1\% level of significance and the result is presented in Tables 1. From Tables 1, we observe that  the empirical type 1 error are close to the chosen level of significance. We calculated the empirical power of the test and  is given in Table 2. From  the Table 2 we observe  that the test has good power in general and the power increases when sample size increases and the value of $a$ deviates from null hypothesis value ($a=1$). We  also observe from Table 2 that  the power of the test decreases as the censoring percentage increases.

\begin{table}[h]
	\caption{Empirical power the test: Exponential distribution }
	
	\begin{tabular}{|l|l|ll|ll|}
		\hline
		$(a,P_1)$& $n$&\multicolumn{2}{l|}{  20\% Censored}& \multicolumn{2}{l|}{  40\% Censored}\\
		\cline{3-6}
		&&  5 \% level &1 \% level&  5 \% level &1 \% level\\
		\hline
		\multirow{6}[10]{*}{$(1.5,0.3)$}
		&	50&0.5412  &0.4652 &0.3404 &0.2863 \\
		&	75&0.6564 &0.5809 &0.3898 &0.3279\\
		&	100&0.8022 &0.6944 &0.5683 &0.5021\\
		&	150&0.9682 &0.8048 &0.7656 &0.6892\\
		&  200&1.0000&0.9620 & 0.8834&0.8238\\
		\hline
		\multirow{6}[10]{*}{$(1.9,0.3)$}
		&	50&0.6802 & 0.5585& 0.4121&0.3782\\
		&	75& 0.7106& 0.6011&0.4983 &0.4387\\
		&	100&0.8604 &0.7240 &0.6482 &0.5741\\
		&	150&0.9805& 0.8436&0.8639&0.7930\\
		&  200&1.0000& 0.9829&0.9422 &0.9058\\
		\hline
		
	\end{tabular}%
	\label{ta1}%
\end{table}%

Next, we simulate lifetime from Weibull random variable with shape parameter  $\theta=2$ where the distribution function is given by ${F}(x)=1-e^{-x^{\lambda}}$,\, $\lambda>1$, $x\geq 0$. The censoring variable $C$ and truncated variable $L$ are simulated as above. The empirical power obtained in this case is  reported in Table 3. In this case also, from Table 3, we observe that power increases as the  sample size increases.

\begin{table}[h]
	\caption{Empirical power the test: Weibull distribution }
	
	\begin{tabular}{|l|l|ll|ll|}
		\hline
		$(a,P_1)$& $n$&\multicolumn{2}{l|}{  20\% Censored}& \multicolumn{2}{l|}{  40\% Censored}\\
		\cline{3-6}
		&&  5 \% level &1 \% level&  5 \% level &1 \% level\\
		\hline
		\multirow{6}[10]{*}{$(1.5,0.3)$}
		&	50&0.5219  &0.4363 &0.3331 &0.2728 \\
		&	75&0.6436 &0.5602 &0.3938 &0.3040\\
		&	100&0.8001 &0.6761 &0.5459 &0.4983\\
		&	150&0.9529 &0.7883 &0.7581 &0.6634\\
		&  200&1.0000&0.9636 & 0.8776&0.8033\\
		\hline
		\multirow{6}[10]{*}{$(1.9,0.3)$}
		&	50&0.6466 & 0.5375& 0.4121&0.3584\\
		&	75& 0.7049& 0.5980&0.4831 &0.4158\\
		&	100&0.8448 &0.7042 &0.6160 &0.5681\\
		&	150&0.9763& 0.8389&0.8445&0.7708\\
		&  200&1.0000& 0.9801&0.9384 &0.9003\\
		\hline
			\end{tabular}%
	\label{ta1}%
\end{table}%

\subsection{Data analysis}

In this section, we illustrate the proposed test procedure using the real data set mentioned in Hong et al. (2009). The data provide the information of lifetime of transformers from an energy company. There were approximately 15000 transforms and the company started recording the information about the transformers in 1980.  For the analysis Hong et al. (2009) considered  the data till 2008. The data contain the information of transformers which are installed before or after 1980 but failed after 1980. The lifetime of the transformers which are  still in service in 2008 is considered as right censored. Moreover, no information was available for the units which installed and failed before 1980.  Also the data contain the information about the possible causes of failure of each unit.  Hence the lifetime of the transformers  can be treated as  left truncated and right censored competing risks data. Many authors considered the analysis of this data. Balakrishnan and Mitra (2012) and Kundu et al. (2017) considered the extract of this data (data of sample size 100) for the analysis. For analysing of transformers data,  Kundu et al. (2017) developed a parametric model  for latent failure times under left truncated and right censored competing risks setup. We consider the same data set for the analysis.  In our study, we are interested to test the independence of lifetime of transformers and associated causes of failure.
\begin{small}
\begin{table}[h]
    \caption{Transformer Data}
    \begin{minipage}{.32\linewidth}
\scalebox{0.85}{
        \begin{tabular}{|l|l|l|l|l|}
        \hline
        S.N.& Year  & Year  &$\nu$ &$K$\\
        & Inst. & Exit&&\\
        \hline
           1 & 1961 & 1996 & 0 & 2 \\
           2 & 1964 & 1985 & 0 & 1 \\
           3 & 1962 & 2007 & 0 & 2 \\
           4 & 1962 & 1986 & 0 & 2 \\
           5 & 1961 & 1992 & 0 & 2 \\
           6 & 1962 & 1987 & 0 & 1 \\
           7 & 1964 & 1993 & 0 & 2 \\
           8 & 1960 & 1984 & 0 & 2 \\
           9 & 1963 & 1997 & 0 & 2 \\
           10 & 1962 & 1995 & 0 & 2 \\
           11 & 1963 & 2008 & 0 & 0 \\
           12 & 1963 & 2000 & 0 & 1 \\
           13 & 1960 & 1981 & 0 & 2 \\
           14 & 1963 & 1984 & 0 & 2 \\
           15 & 1963 & 1993 & 0 & 2 \\
           16 & 1964 & 1992 & 0 & 2 \\
           17 & 1961 & 1981 & 0 & 2 \\
           18 & 1960 & 1995 & 0 & 1 \\
           19 & 1961 & 2008 & 0 & 0 \\
           20 & 1960 & 2002 & 0 & 1 \\
           21 & 1960 & 1988 & 0 & 1 \\
           22 & 1961 & 1993 & 0 & 2 \\
           23 & 1961 & 1990 & 0 & 2 \\
           24 & 1960 & 1986 & 0 & 1 \\
           25 & 1962 & 2008 & 0 & 0 \\
           26 & 1964 & 1982 & 0 & 2 \\
           27 & 1963 & 1984 & 0 & 1 \\
           28 & 1960 & 1987 & 0 & 2 \\
           29 & 1962 & 1996 & 0 & 2 \\
           30 & 1963 & 1994 & 0 & 1 \\
           31 & 1987 & 2008 & 1 & 0 \\
           32 & 1980 & 2008 & 1 & 0 \\
           33 & 1988 & 2008 & 1 & 0 \\
           34 & 1985 & 2008 & 1 & 0 \\
           \hline
             \end{tabular}}
               \end{minipage}%
            \begin{minipage}{.32\linewidth}
            \scalebox{0.85}{
                  \begin{tabular}{|l|l|l|l|l|}
                         \hline
                         S.N.& Year & Year  &$\nu$ &$K$\\
                         & Inst. & Exit&&\\
                         \hline
           35 & 1989 & 2008 & 1 & 0 \\
           36 & 1981 & 2008 & 1 & 0 \\
           37 & 1985 & 2008 & 1 & 0 \\
           38 & 1986 & 2004 & 1 & 2 \\
           39 & 1980 & 1987 & 1 & 2 \\
           40 & 1986 & 2005 & 1 & 1 \\
           41 & 1980 & 2008 & 1 & 0 \\
           42 & 1982 & 2008 & 1 & 0 \\
           43 & 1986 & 2008 & 1 & 0 \\
           44 & 1984 & 2008 & 1 & 0 \\
           45 & 1986 & 1995 & 1 & 2 \\
           46 & 1986 & 2008 & 1 & 0 \\
           47 & 1987 & 2008 & 1 & 0 \\
           48 & 1986 & 2008 & 1 & 0 \\
           49 & 1986 & 2008 & 1 & 0 \\
           50 & 1984 & 2008 & 1 & 0 \\
            51 & 1984 & 2001 & 1 & 2 \\
            52 & 1983 & 2008 & 1 & 0 \\
            53 & 1988 & 2008 & 1 & 0 \\
            54 & 1988 & 2008 & 1 & 0 \\
            55 & 1985 & 2008 & 1 & 0 \\
            56 & 1986 & 2008 & 1 & 0 \\
            57 & 1988 & 2008 & 1 & 0 \\
            58 & 1982 & 2008 & 1 & 0 \\
            59 & 1985 & 2008 & 1 & 0 \\
            60 & 1988 & 2008 & 1 & 0 \\
            61 & 1982 & 2004 & 1 & 2 \\
            62 & 1980 & 2008 & 1 & 0 \\
            63 & 1980 & 2002 & 1 & 2 \\
            64 & 1984 & 2008 & 1 & 0 \\
            65 & 1981 & 1999 & 1 & 1 \\
            66 & 1986 & 2007 & 1 & 2 \\
            67 & 1987 & 2008 & 1 & 0 \\
            68 & 1983 & 2008 & 1 & 0 \\
            \hline
             \end{tabular}}
                \end{minipage}%
             \begin{minipage}{.3\linewidth}
             \scalebox{0.85}{
                   \begin{tabular}{|l|l|l|l|l|}
                          \hline
                          S.N.& Year & Year &$\nu$ &$K$\\
                          & Inst. & Exit&&\\
                          \hline
            69 & 1983 & 2006 & 1 & 2 \\
            70 & 1983 & 1993 & 1 & 1 \\
            71 & 1989 & 2008 & 1 & 0 \\
            72 & 1989 & 2008 & 1 & 0 \\
            73 & 1986 & 2008 & 1 & 0 \\
            74 & 1982 & 1999 & 1 & 2 \\
            75 & 1985 & 2008 & 1 & 0 \\
            76 & 1986 & 2008 & 1 & 0 \\
            77 & 1982 & 2008 & 1 & 0 \\
            78 & 1988 & 2004 & 1 & 1 \\
            79 & 1980 & 2008 & 1 & 0 \\
            80 & 1982 & 2002 & 1 & 2 \\
            81 & 1981 & 2006 & 1 & 2 \\
            82 & 1988 & 1996 & 1 & 1 \\
            83 & 1985 & 2002 & 1 & 2 \\
            84 & 1984 & 2008 & 1 & 0 \\
            85 & 1980 & 2008 & 1 & 0 \\
            86 & 1982 & 2008 & 1 & 0 \\
            87 & 1981 & 1995 & 1 & 2 \\
            88 & 1986 & 1997 & 1 & 2 \\
            89 & 1986 & 2008 & 1 & 0 \\
            90 & 1986 & 2008 & 1 & 0 \\
            91 & 1982 & 2008 & 1 & 0 \\
            92 & 1989 & 2008 & 1 & 0 \\
            93 & 1984 & 2008 & 1 & 0 \\
            94 & 1980 & 2008 & 1 & 0 \\
            95 & 1988 & 2008 & 1 & 0 \\
            96 & 1986 & 2008 & 1 & 0 \\
            97 & 1982 & 1996 & 1 & 2 \\
            98 & 1982 & 2008 & 1 & 0 \\
            99 & 1982 & 2008 & 1 & 0 \\
            100 & 1989 & 2008 & 1 & 0 \\
            &&&&\\
            &&&&\\
            \hline
        \end{tabular}}
    \end{minipage}
\end{table}
\end{small}
\par The transformed data of size 100 is  presented in Table 4. In Table 4, $\nu$ represents the truncation indicator. Here $\nu=1$ specifies that the transformer was installed after 1980 and  $\nu=0$ specifies that the transformer was installed before 1980.  $K=1$ indicates that the failure is due to cause 1 and $K=2$ indicates that the failure is due to cause 2. $K=0$ denoted the right censored observations. Figure 1 displays the plot of estimator of the cumulative incidence function corresponding to cause 1 and cause 2. From Figure 1, it is evident that the chance of failure due to cause 2 is more compared to cause 1 as the lifetime increases. Now we calculated the value of $\hat \Delta$ and is obtained as 88.16, which indicates that the lifetime and causes of failure are dependent.


\begin{figure}[h!]
  \centering
  \includegraphics[width=10cm]{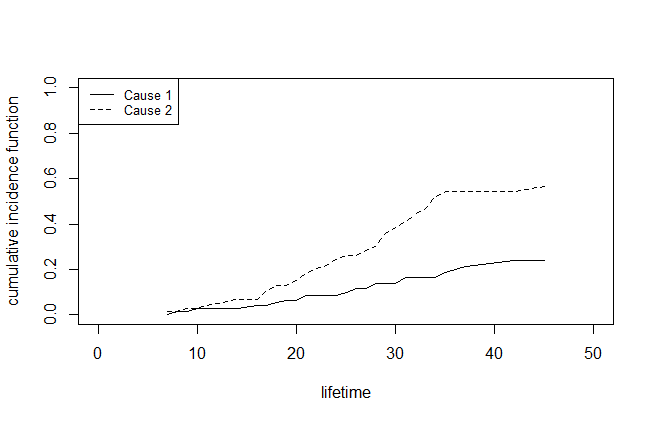}
  \caption{Cumulative incidence functions}
\end{figure}

\section{Conclusion}In many application involving followup studies, lifetime of a patient often subject to  left truncation in addition to the random right censoring. We develop a U-statistics  for left truncated and right censored data.   We use inverse probability weighted technique for developing the U-statistics.  The asymptotic properties of the U-statistics are studied. We proved the $\sqrt{n}$- consistency of the proposed U-statistics. We derived  the asymptotic distribution of the U-statistics  as normal. We then obtained a consistent estimator of the asymptotic variance. As an application, we develop a new test for testing the independence between cause of failure and failure time in competing risks.   The finite ample performance of the test is evaluated through Monte Carlo simulation study.  The test procedure is illustrated using a failure time data of transformers reported by Hong et al. (2009). We established that failure time of the of transformers are dependent on the cause of failure.

  As mentioned in the introduction, a large number of parameters were estimated using U-statistics. Hence it important  to develop U-statistics under  different  censoring scheme.  Some important censoring mechanism  appeared  in medical research are double censoring and interval censoring. Extending our works to these censoring scheme can be considered for future works.

\noindent{\bf \large  Appendix}

\noindent {\bf Proof of Theorem 1:} We prove the theorem for $m=1,2$ and   proofs for the other cases are similar.
Consider
\begin{small}
\begin{eqnarray}\label{consi}
\widehat U_1&=& \frac{1}{n} \sum_{i=1}^{n}\frac{h(T_{i})  \delta_i\epsilon_i }{\widehat K_c(T_i)}\nonumber\\
&=& -\frac{1}{n} \sum_{i=1}^{n}\frac{h(T_{i})  \delta_i\epsilon_i (\widehat K_C(T_i)-P(L_i<T_i<C_i))}{\widehat K_c(T_i)P(L_i<T_i<C_i)}+\frac{1}{n} \sum_{i=1}^{n}\frac{h(T_{i})  \delta_i\epsilon_i }{P(L_i<T_i<C_i)}\nonumber\\&=&A_1+A_2\qquad(\text{say}).
\end{eqnarray}
\end{small}Since $A_2$ is a U-statistic  with kernel $\frac{h(T_{i})  \delta_i\epsilon_i }{P(L_i<T_i<C_i)}$, as $n\rightarrow\infty$, $A_2$ converges in probability to $\theta$. Note that  $\widehat K_c (T_i)$ is a consistent estimator of $P(L_i<T_i<C_i)$. Also we have
\begin{equation*}
 \max_{1\le i\le n} \frac{\sqrt{n} \big| {\widehat K_c (T_i)-P(L_i<T_i<C_i))}\big|}{\widehat K_c (T_i)}=O_p(1),
\end{equation*} Now, consider
\begin{eqnarray*}
  |A_1| &=&\Big|-\frac{1}{n} \sum_{i=1}^{n}\frac{h(T_{i})\delta_i\epsilon_i (\widehat K_C(T_i)-P(L_i<T_i<C_i))}{\widehat K_C(T_i)P(L_i<T_i<C_i)}\Big| \\
   &\le& \begin{small} \max_{1\le i\le n}\frac{\sqrt{n}\Big|(\widehat K_C(T_i)-P(L_i<T_i<C_i)) \Big|}{\widehat K_C(T_i)} \frac{1}{\sqrt{n}}.\frac{1}{n} \sum_{i=1}^{n}\Big|\frac{h(T_{i})\delta_i\epsilon_i }{P(L_i<T_i<C_i)}\Big|\end{small} \\
   &\le&O_p(1)o_p(1)O_p(1)=o_p(1).
\end{eqnarray*}Hence from the representation given in (\ref{consi}), we have the result for $m=1$.

Next, consider the case $m=2$, where $\widehat{U}_c$ is  given by
\begin{eqnarray}\label{rep}
 \widehat{U}_2&=&\frac{2}{n(n-1)}\sum_{i=1}^{n}\sum_{j<i;j=1}^{n}\frac{h(T_i,T_{j})\delta_i\delta_j\epsilon_i\epsilon_j}{\widehat{K}_c(T_i)\widehat{K}_c(T_j)}\nonumber\\
&=&\nonumber\frac{2}{n(n-1)}\sum_{i=1}^{n}\sum_{j<i;j=1}^{n}\frac{h(T_i,T_{j})\delta_i\delta_j(\widehat{K}_c(T_j)-P(L_j<T_j<C_j))}
{\widehat{K}_c(T_i)\widehat{K}_c(T_j)}\nonumber\\&&\nonumber\qquad\qquad\qquad\qquad\qquad\times\frac{(\widehat{K}_c(T_i)-P(L_i<T_i<C_i))}{P(L_i<T_i<C_i)P(L_j<T_j<C_j)}
\\ \nonumber &\quad&+\frac{2}{n(n-1)}\sum_{i=1}^{n}\sum_{j<i;j=1}^{n}\frac{h(T_i,T_{j})\delta_i\delta_j}{\widehat{K}_c(T_i)P(L_j<T_j<C_j)}
\\ \nonumber &\quad&+\frac{2}{n(n-1)}\sum_{i=1}^{n}\sum_{j<i;j=1}^{n}\frac{h(T_i,T_{j})\delta_i\delta_j}{P(L_i<T_i<C_i)\widehat{K}_c(T_j)}
\\ \nonumber &\quad&-\frac{2}{n(n-1)}\sum_{i=1}^{n}\sum_{j<i;j=1}^{n}\frac{h(T_i,T_{j})\delta_i\delta_j}{P(L_i<T_i<C_i)P(L_j<T_j<C_j)}
\\ &=&\widehat U_{12}+\widehat U_{22}+\widehat U_{32}-\widehat U_{42}.\qquad\qquad(\text{say})\end{eqnarray}
Note that $\widehat{K}_c(T_1)$ is a consistent estimator of $P(L_1<T_1<C_1)$.  Hence, as $n\rightarrow \infty$
\begin{small}
\begin{eqnarray}\label{13}
\nonumber |\widehat U_{12}|&\le&\nonumber \sup_{T_i} |(\widehat{K}_c(T_i)-P(L_i<T_i<C_i))|\sup_{T_j} |(\widehat{K}_c(T_j)-P(L_j<T_j<C_j))|\nonumber\\ \nonumber &\quad&\frac{2}{n(n-1)}\sum_{i=1}^{n}\sum_{j<i;j=1}^{n}|\frac{h(T_i,T_{j})\delta_i\delta_j}{\widehat{K}_c(T_i)\widehat{K}_c(T_j)P(L_i<T_i<C_i)P(L_j<T_j<C_j)}|
\\ &=&o_p(1)o_p(1)O_p(1)=o_p(1).\end{eqnarray}\end{small}
Similar lines  as above we can show that
\begin{eqnarray}\label{14}
\widehat U_{22}&=&\widehat U_{42}+\frac{2}{n(n-1)}\sum_{i=1}^{n}\sum_{j<i;j=1}^{n}\frac{h(T_i,T_{j})\delta_i\delta_j(\widehat{K}_c(T_i)-P(L_i<T_i<C_i))}{P(L_i<T_i<C_i)\widehat{K}_c(T_i)P(L_j<T_j<C_j)}
\nonumber \\&=&\widehat U_{42}+o_p(1).\end{eqnarray}
and
\begin{eqnarray}\label{15}
\widehat U_{32}&=&\widehat U_{42}+\frac{2}{n(n-1)}\sum_{i=1}^{n}\sum_{j<i;j=1}^{n}\frac{h(T_i,T_{j})\delta_i\delta_j(\widehat{K}_c(T_j)-P(L_j<T_j<C_j))}{P(L_i<T_i<C_i)\widehat{K}_c(T_j)P(L_j<T_j<C_j)}
\nonumber \\&=&\widehat U_{42}+o_p(1).\end{eqnarray}
 Substituting  equations (\ref{13}), (\ref{14}) and (\ref{15}) in equation (\ref{rep}) we obtain
\begin{eqnarray}\label{repnew}
\nonumber \widehat{U}_2  &=&\widehat U_{42}+o_p(1).\end{eqnarray}
We observe that $\widehat U_{42}$  is a U-statistic  with kernel  $\frac{h(Y_i,Y_{j})\delta_i\delta_j\epsilon_i\epsilon_j}{P(L_i<T_i<C_i)P(L_j<T_j<C_j)}$. Hence, as $n\rightarrow\infty$, $\widehat U_{42}$  converges in probability to $\theta$ (Lehmann, 1951).  Accordingly, for $m=2$, as $n\rightarrow\infty$, $\widehat U_{2}$ converges in probability to $\theta$.

\noindent {\bf Proof of Theorem 2:} For convenience, denote $K_c(T_i)=P(L_i<T_i<C_i)$ for any $i=1,\ldots,n$.  First, we consider  the decomposition
\begin{equation}\label{decomp}
  \sqrt{n}(\widehat{U}_m-\theta)= \sqrt{n}(U_m-\theta)+\sqrt{n}(\widehat{U}_m-U_m).
\end{equation}Since  $\widehat K_c (T_i)$ is a $\sqrt{n}$-consistent for $K_c(T_i)$, we have $\sqrt{n}(\widehat{K}_c(T_i)-K_c(T_i))=O_p(1).$ Hence the second term in the decomposition (\ref{decomp}) can be written as (for the detailed steps see the proof above to show $\widehat U_{12}=o_p(1)$)
\begin{small}
 \begin{eqnarray}\label{decomp2}
  \sqrt{n}(\widehat{U}_m-U_m)&=&\frac{\sqrt{n}}{\dbinom{n}{m}}\sum_{1\le i_1<\ldots<i_m}\frac{h(T_{i_{1}},\ldots,T_{i_{m}})\prod_{l\in \underline{i}}\delta_{l}\epsilon_l\Big(\prod_{l\in \underline{i}}(\widehat K_c(T_l)-K_c(T_l))\Big)}{\prod_{l\in \underline{i}}K_c(T_l)\prod_{l\in \underline{i}}K_c(T_l)}\nonumber \\&&\qquad\qquad\qquad\qquad+o_p(1).\nonumber
\end{eqnarray}
\end{small}
Hence using Hoeffding (1948) decomposition, from equation (\ref{decomp}) we have
\begin{eqnarray}\label{eq26}
\nonumber  \sqrt{n}(\widehat{U}_m-\theta)&=& \frac{m}{\sqrt n}\sum_{i=1}^{n}\frac{h_1(T_i)\delta_i\epsilon_i}{K_c(T_i)}-\theta\\&-&\frac{m}{\sqrt n}\sum_{i=1}^{n}\frac{h_1(T_i)\delta_i\epsilon_i(\widehat{K}_c(T_i)-{K}_c(T_i))}{K_{c}^2(T_i)}+o_p(1),
\end{eqnarray}
where
\begin{equation*}
 h_{1}(x)=E\big(h(T_{1},...,T_{m})|T_1=x\big).
\end{equation*}
 Using the relationship  between $\widehat{K}_c(.)$  and $\widehat{\Lambda}_c(.)$  given in (\ref{kmrel}) and by delta method we have
\begin{eqnarray*}
 \sqrt{n}(\widehat{K}_c(T_i)-K_c(T_i))&=&-\sqrt{n}K_c(T_i)(\widehat{\Lambda}_c(T_i)-{\Lambda}_c(T_i))+o_p(1).
\end{eqnarray*}Hence using the martingale representation of $\widehat{\Lambda}_c(T_i)$ (see Page 178 of Andersen et al. (1993)), equation  (\ref{eq26}) becomes
\begin{eqnarray}\label{eq27}
\nonumber  \sqrt{n}(\widehat{U}_m-\theta)&=& \frac{m}{\sqrt n}\sum_{i=1}^{n}\frac{h_1(T_i)\delta_i\epsilon_i}{K_c(T_i)}-\theta\\&&+\frac{m}{\sqrt n}\sum_{i=1}^{n}\frac{h_1(T_i)\delta_i\epsilon_i(\widehat{\Lambda}_c(T_i)-{\Lambda}_c(T_i))}{K_c(T_i)}+o_p(1)\nonumber \\&=& \nonumber\frac{m}{\sqrt n}\sum_{i=1}^{n}\frac{h_1(T_i)\delta_i\epsilon_i}{K_c(T_i)}-\theta\\
&&+\frac{m}{\sqrt n}\sum_{i=1}^{n}\frac{h_1(T_i)\delta_i\epsilon_i}{K_c(T_i)}\int_{0}^{T_i}\frac{d{M}^c(x)}{Y(x)}+o_p(1),
\end{eqnarray}
provided $Y(x)>0$ with probability one.

  Now, recall the definition of $S(x)$ given in equation (\ref{subdist}),  we can express 
\begin{small}
\begin{equation*}
m \sqrt n\frac{1}{n}\sum_{i=1}^{n}\frac{h_1(T_i)\delta_i\epsilon_i}{K_c(T_i)}\int_{0}^{T_i}\frac{d{M}^c(x)}{Y(x)}={m}{\sqrt n}\int_{0}^{\infty}\frac{h_1(z)}{K_c(z)}\left(\int_{0}^{z}\frac{d{M}^c(x)}{Y(x)}\right)dS(z).
\end{equation*}
\end{small}
Using Fubini's theorem, changing the order of integration gives
\begin{small}
\begin{equation*}
m \sqrt n\frac{1}{ n}\sum_{i=1}^{n}\frac{h_1(T_i)\delta_i\epsilon_i}{K_c(T_i)}\int_{0}^{T_i}\frac{d{M}^c(x)}{Y(x)}= m \sqrt n \int_{0}^{\infty}\left(\int_{x}^{\infty}\frac{h_1(z)}{P(L<z<C)}dS(z)\right)\frac{d{M}^c(x)}{Y(x)}.
\end{equation*}
\end{small}We denote the right hand side of the above equation as
\begin{small}
\begin{equation*}
m \sqrt n\int_{0}^{\infty}\frac{w(x)}{Y(x)}d{M}^c(x),
\end{equation*}
\end{small}where $w(x)=\int_{0}^{\infty}\frac{h_1(z)I(z>x)}{K_c(z)}dS(z)$. Hence we can write  $(\ref{eq27})$ as
\begin{eqnarray*}\label{decomp1}
\nonumber  \sqrt{n}(\widehat{U}_m-\theta)&=& \frac{m}{\sqrt n}\sum_{i=1}^{n}\frac{h_1(T_i)\delta_i\epsilon_i}{K_c(T_i)}-\theta+{m}{\sqrt n}\int_{0}^{\infty}\frac{w(x)}{Y(x)}d{M}^c(x)+o_p(1)\nonumber\\
&=&D_1+D_2+o_p(1). \qquad\text{(say)}
\end{eqnarray*}
Using central limit theorem, as $n\rightarrow\infty$, $D_1$ converges in distribution to Gaussian random variable with mean zero and variance $\sigma_{1}^{2}$ where $\sigma_{1}^{2}$ is given by
\begin{equation*}
  \sigma_{1}^{2}=m^2Var\Big(\frac{h_1(X)\delta_1\epsilon_1}{K_c(X)}\Big).
\end{equation*}
As $n\rightarrow\infty$, $Y(x)/n$ converges in probability to $y(x)$. Using martingale central limit theorem,  as $n\rightarrow\infty$, $D_2$ converges in distribution to Gaussian random variable with mean zero and variance $\sigma_{2}^2$, where $\sigma_{2}^{2}$ is the limit of the predictable variation  process  given by
\begin{eqnarray*}
  \sigma_{2}^{2}&=& \lim_{n\rightarrow\infty}{m^2}n \int_{0}^{\infty}\frac{w^2(x)Y(x)d{\Lambda}_c(x)}{Y^2(x)}\\
  &=& \lim_{n\rightarrow\infty}{m^2} \int_{0}^{\infty}\frac{w^2(x)d{\Lambda}_c(x)}{Y(x)/n}
  \\
  &=& {m^2} \int_{0}^{\infty}\frac{w^2(x)d{\Lambda}_c(x)}{y(x)}.
\end{eqnarray*}Hence we have the variance expression given in the Theorem 2.2.  The proof is completed when we show that the asymptotic covariance between $D_1$ and $D_2$ is zero.

Consider
\begin{eqnarray*}
\lefteqn{\hskip -2.8cm  \Big|\big(\frac{m}{\sqrt n}\sum_{i=1}^{n}\frac{h_1(T_i)\delta_i\epsilon_i}{K_c(T_i)}-\theta\big)\big(m \sqrt n\int_{0}^{\infty}\frac{w(x)}{Y(x)}d{M}^c(x)\big)\Big|} \\ &\hskip -4.8cm =&\hskip -1.8cm \Big|\frac{m}{n}\sum_{i=1}^{n}\frac{h_1(T_i)\delta_i\epsilon_i}{K_c(T_i)}-\theta\Big|\Big|m  n\int_{0}^{\infty}\frac{w(x)}{Y(x)}d{M}^c(x)\Big|
\\ &\hskip -4.8cm \le&\hskip -1.8cm \frac{m}{n}\sum_{i=1}^{n}\Big|\frac{h_1(T_i)\delta_i\epsilon_i}{K_c(T_i)}-\theta\Big|\Big|m  \int_{0}^{\infty}\frac{w(x)}{Y(x)/n}d{M}^c(x)\Big|\\ &\hskip -4.8cm \le&\hskip -1.8cm O_p(1).o_p(1)=o_p(1).
\end{eqnarray*}
This completes the proof of the theorem.


\begin{thebibliography}{xx}


\bibitem{} Andersen, P. K., Borgan, O., Gill, R. D. and Keiding, N. (1993). {\em Statistical models based on counting processes.} Springer Science \& Business Media, New York.

\bibitem{} Anjana S, Isha Dewan and Sudheesh, K. K. (2019). Test for independence between time to failure and cause of failure in competing risks with $k$ causes of failure. {\em Journal of Nonparametric Statistics,} 31, 322--339.

\bibitem{} Balakrishnan, N., Mitra, D. (2011). Likelihood inference for lognormal data with left truncation and right censoring with an illustration. {\em Journal of Statistical
Planning and Inference}, 141, 3536--3553.

\bibitem{} Balakrishnan, N. and Mitra, D. (2012). Left truncated and right censored Weibull data and likelihood inference with an illustration. {\em Computational Statistics and Data Analysis}, 56,4011--4025.


\bibitem{} Balakrishnan, N. and  Mitra, D. (2014). Some further issues concerning likelihood inference for left truncated and right censored lognormal data. {\em Communications in Statistics-Simulation and Computation}, 43, 400--416.


 \bibitem{} Chen, Y., and Datta, S. (2019). Adjustments of multi-sample U-statistics to right censored data and confounding covariates. {\em Computational Statistics \& Data Analysis, } 135, 1--14.

\bibitem{}  Chen, C. M., and  Shen, P. S. (2018). Conditional maximum likelihood estimation in semiparametric transformation model with LTRC data. {\em Lifetime Data Analysis}, 24, 250--272.

\bibitem{} Chen, C. M., Shen, P. S., Wei, J. C. C.,  and  Lin, L. (2017). A semiparametric mixture cure survival model for left truncated and right censored data. {\em Biometrical Journal,} 59, 270--290.

\bibitem{} Chen, L. P. and Yi, G. Y. (2021). Semiparametric methods for left-truncated and right-censored survival data with covariate measurement error. {\em Annals of the Institute of Statistical Mathematics,} 73, 481--517.

\bibitem{} Crowder, M. J. (2012). {\em Multivariate Survival Analysis and Competing Risks.} CRC Press, Boca Raton.


\bibitem{}  Hoeffding, W. (1948).
A class of statistics with asymptotically normal distribution. {\em Annals of Mathematical Statistics}, 19,  293--325. 

\bibitem{}
Cortese, G., Holmboe, S. A., and  Scheike, T. H. (2017). Regression models for the restricted residual mean life for right censored and left truncated data. {\em Statistics in Medicine,}  36, 1803--1822.


\bibitem{}
Datta, S., Bandyopadhyay, D., and Satten, G. A. (2010). Inverse Probability of Censoring Weighted U-statistics for Right Censored Data with an Application to Testing Hypotheses. {\em Scandinavian Journal of Statistics,} 37, 680--700.

\bibitem{} Dewan, I., Deshpande, J., and Kulathinal, S. (2004). On testing dependence between time
to failure and cause of failure via conditional probabilities. {\em Scandinavian Journal of
Statistics,} 31, 79--91.

\bibitem{}Dewan, I., and Kulathinal, S. (2007). On testing dependence between time to failure and cause of failure when causes of failure are missing. {\em PloS One}, 2, 1255--1264.

\bibitem{} Efromovich, S. and Chu, J. (2018). Hazard rate estimation for left truncated and right censored data. {\em Annals of the Institute of Statistical Mathematics}, 70, 889--917.

\bibitem{} Geskus, R. B. (2011). Cause-specific cumulative incidence estimation and the fine and gray model under both left truncation and right censoring. {\em Biometrics}, 67, 39--49.


\bibitem{} Hong, Y., Meeker, W. Q., and  McCalley, J. D. (2009). Prediction of remaining life of power transformers based on left truncated and right censored lifetime data. {\em The Annals of Applied Statistics}, 3, 857--879.


\bibitem{} Hou, J., Chambers, C. D., and  Xu, R. (2018). A nonparametric maximum likelihood approach for survival data with observed cured subjects, left truncation and right-censoring. {\em Lifetime Data Analysis}, 24, 612--651.



\bibitem{} Jiang, H., Fine, J. P., and  Chappell, R. (2005). Semiparametric analysis of survival data with left truncation and dependent right censoring. {\em Biometrics,} 61, 567--575.

\bibitem{} Jiang, W., Ye, Z. and Zhao, X. (2020). Reliability estimation from left-truncated and right-censored data using splines. {\em Statistica Sinica,} 30, 845--875.
%
 \bibitem{} Jing, B. Y., Yuan, J. and Zhou, W. (2009). Jackknife empirical likelihood. {\it Journal of the American Statistical Association}, 104,
  1224--1232.


\bibitem{} Kalbfeisch, J. D., and Prentice, R. L. (2002). {\em The Statistical Analysis of Failure Time Data.}
John Wiley \& Sons, New York.


\bibitem{} Klein, J. P., and Moeschberger, M. L. (2006). {\em Survival Analysis: Techniques for Censored and Truncated Data}. Springer Science \& Business Media, New York.



\bibitem{} Kowalski, J.  and Xin M. T. (2008).  {\em Modern Applied U-Statistics.} John Wiley \& Sons, New Jersey.

\bibitem{} Kundu, D., Mitra, D., and Ganguly, A. (2017). Analysis of left truncated and right censored competing risks data. {\em Computational Statistics \& Data Analysis}, 108, 12--26.

\bibitem{} Lawless, J. F. (2003). {\em Statistical Models and Methods for Lifetime Data}. John Wiley \&
Sons, New York.





\bibitem{} Lee, A. J. (1990). {\em U-Statistics: Theory and Practice}.  CRC Press, Boca Raton.


\bibitem{} Lehmann, E. L. (1951), Consistency and unbiasedness of certain nonparametric tests.
{\em  The Annals of Mathematical Statistics}, 22, 165--179.


\bibitem{} Satten, G. A., Kong, M., and Datta, S. (2018). Multisample adjusted U-statistics that account for confounding covariates. {\em Statistics in Medicine}, 37, 3357--3372.



%
%
%



\bibitem{} Su, Y. R. and Wang, J. L. (2012). Modeling left-truncated and right-censored survival data with longitudinal covariates. {\em The Annals of Statistics}, 40, 1465--1488.

\bibitem{} \vskip8pt\noindent  Vakulenko‐Lagun, B.,and  Mandel, M. (2016). Comparing estimation approaches for the illness–death model under left truncation and right censoring. {\em Statistics in Medicine}, 35, 1533--1548.

\bibitem{} Zhang, X., Zhang, M. J.,  and  Fine, J. (2011). A proportional hazards regression model for the subdistribution with right censored and left truncated competing risks data. {\em Statistics in Medicine}, 30, 1933--1951.
\end{thebibliography}
 \end{document}